\documentclass[preprint,12pt]{elsarticle}

 \usepackage{graphicx}
\usepackage{amssymb}
 \def \g{\gamma}
 \def \G{\Gamma}
\def \d{\partial}
 \def \o{\omega}
  
  \def \b{\beta}

 \def \e{\varepsilon}
 
 \def\be{\begin{equation}}
 \def\ee{\end{equation}}
 
\journal{Optics Communications}

\begin{document}

\begin{frontmatter}



%\title{Spatial dispersion of nanocavity plasmonic polaritons}
\title{Confined plasmonic modes in a nanocavity}

\author{Aurore Castani\'e and Didier Felbacq}

\address{Université de Montpellier 2, Laboratoire Charles Coulomb UMR CNRS-UM2 5221\\
Bat.21 CC074, Place Bataillon, 34095 Montpellier Cedex 05, France}

\begin{abstract}
The effect of confinement on surface plasmon polariton in a planar nanocavity was studied. The generalized modes were obtained and studied in detail. It was demonstrated that these modes result from the strong coupling of plasmon-like and photon-like modes.
\end{abstract}


\end{frontmatter}

% \linenumbers

\section{Introduction}
\label{}
Surface Plasmon Polaritons are surface mode localized at the interface between a dielectric and a metal that can be confined in regions much smaller than their wavelength. Because of this strong enhancement many interesting phenomena are linked to plasmons: a huge Purcell effect \cite{PRL}, the strong coupling with excitons \cite{bellessa}, quantum dots luminescence \cite{QD}, spasers \cite{stockman}. All this has resulted into the new field of Plasmonics \cite{chalaev}. It has even be suggested that plasmons could allow for a control of light at the quantum level \cite{chang}.

In this work, we study the effect of confinement on surface plasmons in a planar cavity \cite{Atwater} with a wall coated with a lossy metal. The dispersion curves are studied in details and explicit relations are obtained locally. In particular, it is shown that the modes result from the strong coupling of photon-like modes with plasmon-like modes. It is shown that modes strongly concentrated at the air-metal interface can be obtained. Finally, we discuss the modes of the lossy waveguide, in particular their completeness.

\begin{figure}
   \begin{center}
 \begin{tabular}{c}
   \includegraphics[width=8cm]{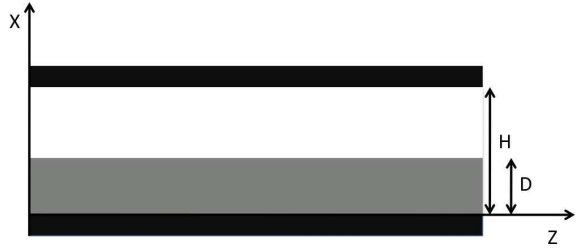}
  \end{tabular}
   \end{center}
   \caption[example] 
   { \label{cavity}The structure under study.}
   \end{figure} 
 \section{The dispersion relation}
 The structure under study is depicted in fig. (\ref{cavity}). We consider $H_{||}$ modes only, that is electromagnetic fields for which the magnetic field is linearly polarized along the $z$ direction. Time harmonic field with a time dependence of $e^{-ik_0 c t}$ are considered ($c$ is the speed of light in vacuum). We denote $H(x,y)=u(x) e^{i\g y}\, e^{-ik_0 c t}\, e_z$. The total heigth of the waveguide is $h$ and the thickness of the metal is $d$. We denote $\tau=d/h$. The field $u(x)$ satisfies the following equation, in the Schwartz distributions meaning:
 \be \label{spectral}
 \d_x \left(\e^{-1} \d_x u\right)+\left(k_0^2 -\g^2 \e^{-1} \right)u=0
 \ee
 where $\e(x)$ is equal to $1$ for $x \in [d,h]$ and to $\e_m=1-\frac{\o_p^2}{\o(\o+i\Gamma)}$ for $x \in [0,d]$, that is, the metal is described by a Drude model. For numerical computations, the parameters of silver will be used (the plasma wavelength is $137 nm$).
 As a simple model of confinement, we use the Neumann conditions $\partial_x u=0$ on the boundaries of the cavity ($x=0,h$).

 The plasma frequency $\o_p$ allows to define a unit of length by considering the associated wavelength $\lambda=\frac{2\pi c}{\o_p}$ and a unit of spatial frequency by considering the corresponding wavenumber $k_p=\frac{2\pi}{\lambda_p}$. We denote with the upper case the normalized length quantities: $H=k_p h$ the normalized height of the waveguide, $D=k_p d$ the normalized width of the metal layer, $X=k_p x$ the normalized variable, $K_0=k_0/k_p$, $G=\g/k_p$ the normalized wavenumber and propagation constant respectively. We denote
 $$
 \tilde{\beta}_g^2=k_0^2 -\g^2\, , \,
 \tilde{\beta}_m^2=k_0^2 \e_m -\g^2
 $$
 and the corresponding dimensionless quantities:
 \be
 \beta_g^2=K_0^2 -G^2\, , \,
 \beta_m^2=K_0^2 \e_m -G^2
 \ee

A straightforward computation leads to the following expression for the field:
\begin{eqnarray*}
0<x<d: u(x)=B \cos(\beta_m X)  \\
d<x<h: u(x)=C \cos(\beta_g (X-H))
\end{eqnarray*}
The transmission conditions at the interface $X=D$, leads to:
\begin{eqnarray*}
B \cos(\beta_m D)=C \cos(\beta_g (D-H)) \\
B \frac{\beta_m}{\e_m} \sin(\beta_m D)=C \beta_g \sin(\beta_g (D-H))
\end{eqnarray*}
which implies the following dispersion relations:
\begin{equation}\label{eqdisper}
F(K_0^2,G^2)=\frac{\beta_m}{\e_m} \tan(\beta_m H \tau)-\beta_g \tan(\beta_g H(\tau-1))=0
\end{equation}

%Dispersion relation for the plasmon($\tau$ small):
%$$
%(\frac{\beta^4_m}{\e_m}\tau^3-\beta_g^4 -(1-\tau)^3)) h^2/3 + (\frac{\beta^2_m}{\e_m}-\beta^2_g ) \tau+ \beta^2_g =0
%$$
%$$
%\b_g^3( \tan(\b_g h)^2 + 1)\tan(\b_g h) h^2 t^2 + (\frac{\b_m^2}{\e_m} - \b_g^2 (\tan(\b_g\, h)^2 + 1))h t + \b_g\, \tan(\b_g h)=0
%$$
%this shows that with an error of order $O(t^3)$ the dispersion curve is the reunion of
%the  $0th$ order dispersion relation : $k_0^2=\g^2$ the dispersion of vacuum and $\sin(\b_g h)=0$ that of the perfect waveguide
%and 
%to first order:
% $\frac{\beta^2_m}{\e_m}=\beta^2_g $
% which the dispersion relation of the plasmon
% 
% Bare plasmon: $\b_g=-\frac{\b_m}{\e_m}$
% 
% 
%$$
%\tau*(\b_g^2 lp - (\b_m (\b_m lp \tan(\b_m lp)^2 + \b_m lp))./\e_m) + (\b_m \tan(\b_m lp))./\e_m;
%$$
%		
%$$
%\tau h(\b_m^2/\e_m  - \b_g^2 \frac{1}{\cos(\b_g h)^2}) + \b_g  \tan(\b_g h);
%$$
%where is the plasmon? assume $\b_m/\e_m=-\b_g$:
%$$
%\left[-\b_m   -  \frac{\b_g}{\cos(\b_g h)^2}\right] \tau h \b_g+ \b_g  \tan(\b_g h);
%$$
\section{Asymptotics of the dispersion relation}
\begin{figure}
   \begin{center}
 \begin{tabular}{c}
   \includegraphics[width=8cm]{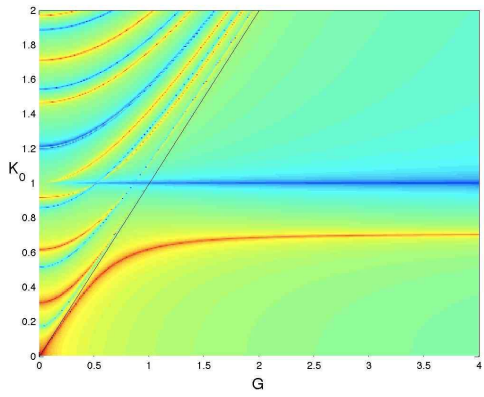}
  \end{tabular}
   \end{center}
   \caption[example] 
   { \label{disper1}Dispersion relation of the waveguide for $\tau=1/5$. The modes appear as yellow-red lines.}
   \end{figure}

In order to give a specific example of dispersion curves, we choose the following parameters:
$h=250nm\,, d=50nm$ (i.e. $\tau=1/5$). A colorplot of $F(K_0^2,G^2)=0$ is given in fig. \ref{disper1}.
Two set of curves can be distinguished: a lower branch, below the plasma frequency of the metal (i.e. $K_0<1$) and below the light cone ($K_0=G$); and an upper branch made of several curves, above the plasma frequency and above the light cone. The curves above the light cone correspond to the usual guided modes, while the mode below the light cone correspond to fields localized at the air-metal interface.

In order to be more specific, we proceed to an asymptotic study of these curves.
Near the ``$\Gamma$ point'': $(k_0,\g)=(0,0)$ and for a waveguide which is small with respect to $\lambda_p$, that is $H\ll 1$, we get the following dispersion relation, from eq. (\ref{eqdisper}):
\be
\b_m^2/\e_m\tau-\b_g^2 (\tau-1)=0
\ee
upon replacing the expression of the various quantities in terms of the principal variables: $K_0$ and $G$, we obtain the relation:
$$
G^2=K_0^2\left(1-\frac{\tau}{K_0^2 +i \G K_0 - 1+\tau}\right)
$$
This equation can be solved for $K_0$ exactly if we assume $\G=0$.
\begin{figure}
   \begin{center}
 \begin{tabular}{c}
   \includegraphics[width=8cm]{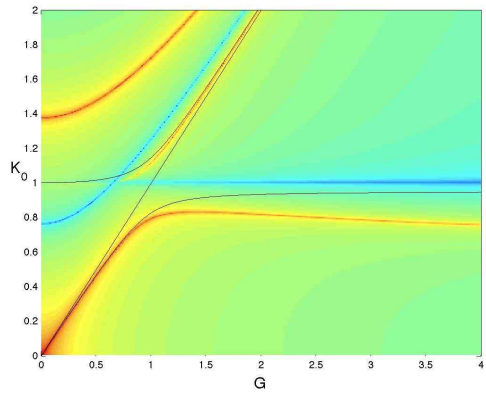}
  \end{tabular}
   \end{center}
   \caption[example] 
   { \label{asdisper1}Dispersion relation of the waveguide for $\tau=1/10$ and $h=50nm$. The modes appear as yellow-red lines. The black lines correspond to the asymptotic dispersion curves $F_u$ and $F_d$.}
   \end{figure} 

We then obtain two branches:
\begin{eqnarray}\label{apdis}
K_0=\frac{\sqrt{2}}{2} \sqrt{1+G^2+ \sqrt{4\, G^2\, \tau + (G^2 - 1)^2}}=F_u(G) \\
K_0=\frac{\sqrt{2}}{2} \sqrt{1+G^2 - \sqrt{4\, G^2\, \tau + (G^2 - 1)^2}}=F_d(G)
\end{eqnarray}
The limits of the branches $F_u$ and $F_d$ are as follows.
As $G$ tends to infinity, $F_d(G)$ tends to the horizontal line $K_0=\sqrt{1 -  \tau}$. On the other hand, when $G \sim 0$, we have: $F_d(G) \sim \sqrt{1-\tau} \, G$.
 \begin{figure}
   \begin{center}
 \begin{tabular}{c}
   \includegraphics[width=8cm]{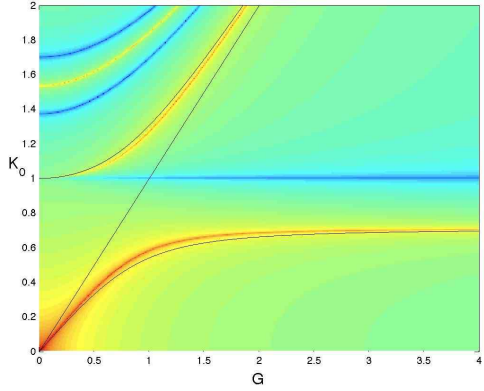}
  \end{tabular}
   \end{center}
   \caption[example] 
   { \label{asdisper2}Dispersion relation of the waveguide for $\tau=1/2$ and $h=50nm$. The modes appear as yellow-red lines. The black lines correspond to the asymptotic dispersion curves $F_u$ and $F_d$.}
   \end{figure}  
For the superior branch, when $G\sim 0$, it holds $F_u(G)\sim 1+\frac{\tau}{2}\,{G}^{2}$, showing that the dispersion relation is parabolic, whereas when $G$ tends to infinity: $F_u(G) \sim G$ and the upper curve is asymptotic to the dispersion of the vacuum.

\begin{figure}
   \begin{center}
 \begin{tabular}{c}
   \includegraphics[width=8cm]{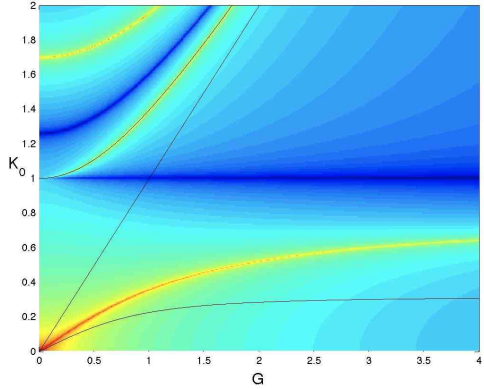}
  \end{tabular}
   \end{center}
   \caption[example] 
   { \label{asdisper3}Dispersion relation of the waveguide for $\tau=9/10$ and $h=50nm$. The modes appear as yellow-red lines. The black lines correspond to the asymptotic dispersion curves $F_u$ and $F_d$.}
   \end{figure}

In fig. (\ref{asdisper1}),(\ref{asdisper2}) and (\ref{asdisper3}), the two branches corresponding to $F_u$ and $F_d$ are given, as well as the colorplot of the dispersion relation. The parameters are $h=50nm$ and $\tau=1/10,\, 1/2,\, 9/10$.
The upper branch $F_u$ is a good approximation for the fundamental (non-plasmonic) mode of the waveguide, while the lower branch is essentially valid near $G=0$. However, the filling ratio $\tau=1/2$ corresponds to a situation where both the upper and lower curves approximate very well the true dispersion curves.
\begin{figure}
   \begin{center}
 \begin{tabular}{c}
   \includegraphics[width=8cm]{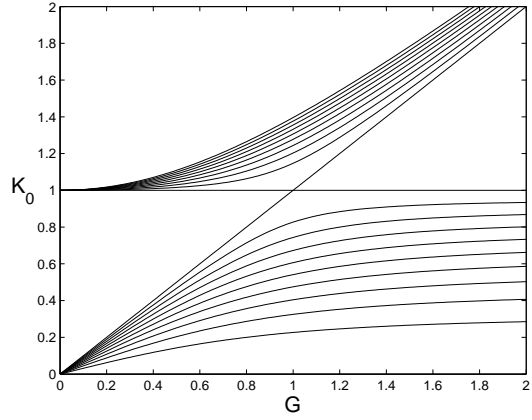}
  \end{tabular}
   \end{center}
   \caption[example] 
   { \label{stroco} Bundle of the approximate dispersion curves given in (\ref{apdis}) for a varying $\tau$ between $0$ and $1$.}
   \end{figure}

 As $\tau$ varies in $[0,1]$, the variation of the curves is given in fig. \ref{stroco}. The part of the curve above $1$ corresponds to those frequencies above the plasma frequency $\o_p$, in which case the real part of the permittivity of the metal is positive. When losses are small enough quasi-modes exist in the guide, leading to the above branch of the dispersion curve. The branch is thus the photon-like curve. The lower branch corresponds to the frequencies below the plasma frequency, for which the real part of the permittivity of the metal is negative. It corresponds to modes essentially localized at the interface of the metal with air, that is, surface plasmons. The lower branch is therefore the plasmon-like curve. When $\tau$ tends to $0$, the tangent dispersion relation becomes:
\be \label{tang1}
K_0=1,K_0=G
\ee
and when $\tau$ tends to $1$:
\be
K_0=0,K_0=\sqrt{G^2+1}
\ee
In particular, the limit case $\tau=0$ shows that the dispersion curve at finite $\tau$ represents an anti-crossing of the two straight lines in eq. (\ref{tang1}), which indicates the strong coupling between the photon modes and the plasmon modes.

\begin{figure}
   \begin{center}
 \begin{tabular}{c}
   \includegraphics[width=8cm]{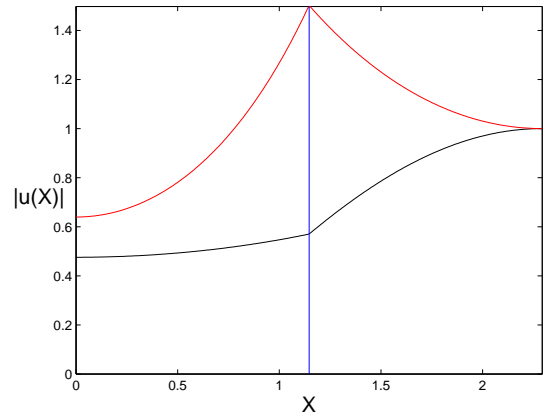}
  \end{tabular}
   \end{center}
   \caption[example] 
   { \label{mode} Modulus of the upper (black) and lower (red) modes at $G=1$. The vertical blue line indicates the air-metal interface.}
   \end{figure}

When $\tau=1/2$, the functions $F_u$ and $F_d$ allows to obtain explicit expressions of the modes with a good approximation.
In fig.\ref{mode}, we have plotted $u(x)$ for $G=1$ and for each branches. The red curve corresponds to the plasmon-like mode and the black one to the photon-like one. It is seen that the red curve has the feature of a field localized at the metal-air interface. 
\begin{figure}
   \begin{center}
 \begin{tabular}{c}
   \includegraphics[width=8cm]{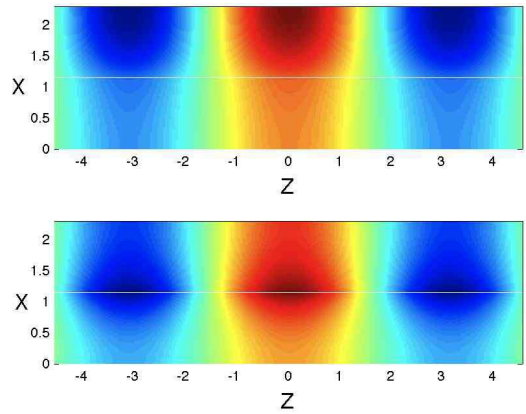}
  \end{tabular}
   \end{center}
   \caption[example] 
   { \label{cartemode} Map of the modes for the upper and lower branches at $G=1$. The modulus is given in fig. (\ref{mode}).}
   \end{figure}
In fig. \ref{cartemode}, the corresponding maps of the field $\Re(u(X)e^{iGZ})$ are given. The upper map corresponds to the photon mode and the lower one to the plasmon mode. The concentration of the field near the metal-air interface is clearly seen.
\begin{figure}
   \begin{center}
 \begin{tabular}{c}
   \includegraphics[width=8cm]{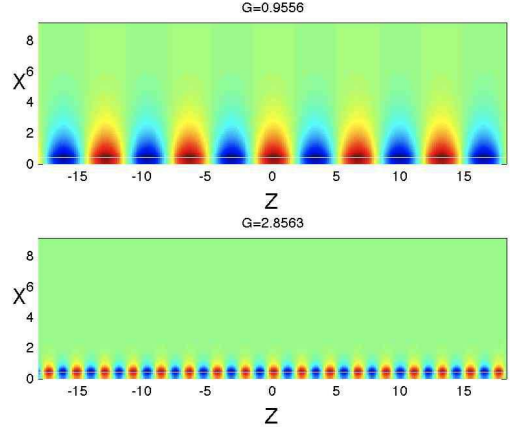}
  \end{tabular}
   \end{center}
   \caption[example] 
   { \label{disdble} Dispersion curves for $\tau=0.05$ and $h=200nm$. The curvature of the plasmonic branch can be seen. }
   \end{figure}
Finally, an interesting situation arises for small filling ratios. For instance for $\tau=1/10$, it can be seen in fig. (\ref{asdisper1}) that the plasmonic branch is curved, in such a way that for some value of $K_0$ (near $0.8$) there are two corresponding values of $G$ satisfying the dispersion relation. Therefore by exciting both modes at their eigenfrequency, one obtains a spatial oscillation between these two modes. Indeed, denoting $\psi_1$ and $\psi_2$ these modes, at the corresponding frequency, a general mode is a linear combination of $\psi_1$ and $\psi_2$ in the form:
$$
\psi(X,Z)=\alpha\, \psi_1(X) e^{iG_1 Z}+\beta\, \psi_2(X) e^{iG_2 Z}
$$
which in modulus takes the usual form of the beating phenomenon:
$$
|\psi(X,Z)|^2=|\alpha\, \psi_1(X)|^2 +|\beta\, \psi_2(X)|^2 +2 \alpha \beta\, \psi_1(X) \psi_2(X) \cos\left((G_1-G_2) Z\right)
$$
Specifically, we choose $K_0=0.72$, solve for $F(K_0^2,G^2)=0$ and find $G_1\sim 0.9556$\,, $G_2\sim 2.8563$. The modulus of the modes is given in fig.(\ref{dblmod1}) and the map of the modes are given in fig.(\ref{dblmod2}). It can be seen on these figures that the smaller spatial period is also much more concentrated at the air-metal interface than the other one. This shows that, despite the fact that the layer of metal is very thin and that the metal is near the plasma frequency, it is still possible to obtain plasmon-like modes strongly localized near the air-metal interface.
\begin{figure}
   \begin{center}
 \begin{tabular}{c}
   \includegraphics[width=8cm]{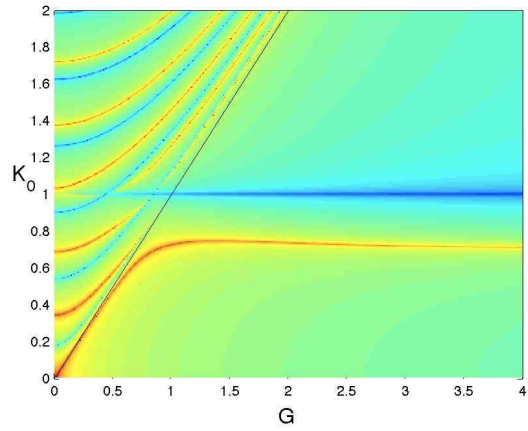}
  \end{tabular}
   \end{center}
   \caption[example] 
   { \label{dblmod1} Modulus of the two plasmonic modes obtained at $K=0.72$ cf. fig. (\ref{disdble}). }
   \end{figure}
\begin{figure}
   \begin{center}
 \begin{tabular}{c}
   \includegraphics[width=8cm]{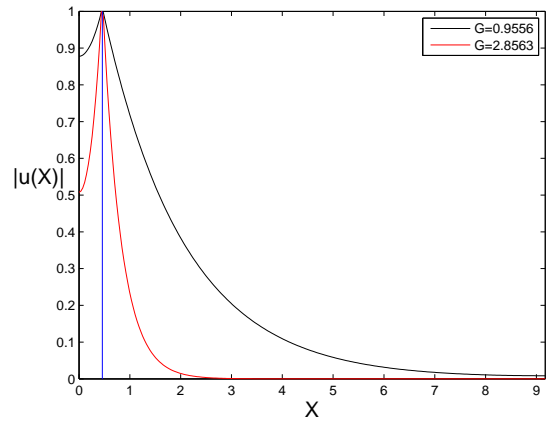}
  \end{tabular}
   \end{center}
   \caption[example] 
   { \label{dblmod2}Map of the two plasmonic modes whose modulus is given in fig. (\ref{dblmod1}).}
   \end{figure}
 The fact that for a small enough filling ratio $\tau$ the plasmonic branch is not monotonic with $G$, i.e. has a maximum and bends is by itself an interesting phenomenon that seems to indicate that the dispersion band can be described by an effective, spatially dispersive, medium. It should be noted that the extreme thinness of the layer of metal questions the validity of the very simple model used here. In fact, such a thin layer is strongly corrugated and the use of a Drude model is questionnable. A more complex model is needed, for instance by describing the surface of the metal as a random rough surface \cite{bra1,bra2,bra3}, to draw sound conclusions. Work is in progress in that direction.
% The   
% Let us plot the set of curves when $\tau$ varies
% 
% 
% Let us write: $-\partial_x(\e^{-1}\partial_x u)+\e^{-1}\g^2 u=k_0^2 u$ and integrate
% 
% The Lagrangian of the electromagnetic field reads as:
% $|E|^2-|B|^2$
% $$
% \int \left(-\e^{-1} |\partial_x u|^2 + \beta^2 |u|^2\right) dx=0
% $$
\section{Some remarks on the modes}
When there are no losses in the layer of metal, the spectral problem defined by equation (\ref{spectral}) provides a complete set of eigenvectors parametrized by $G$.
When $\tau=0$ (i.e. no metal), this corresponds to the following set:
$
\psi_n(X)=\sqrt{\frac{2}{H}}\cos(\frac{n\pi X}{H}), K_0=\sqrt{G^2+ \frac{n^2\pi^2}{H^2}}
$
with the inner product: $(f,g)=\int_0^H f(X)g(X) dX$.

\begin{figure}
   \begin{center}
 \begin{tabular}{c}
   \includegraphics[width=8cm]{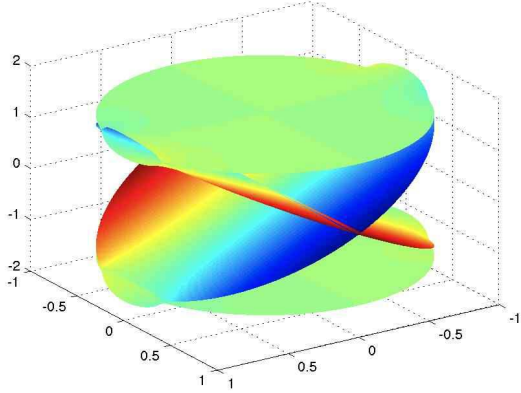}
  \end{tabular}
   \end{center}
   \caption[example] 
   { \label{riem} Four-sheeted Riemann surface corresponding to the approximate dispersion relations given in eq. (\ref{apdis}).}
   \end{figure}

The adjunction of losses implies that the solutions to $F(K_0^2,G^2)=0$ are no longer real.
 However, it is still possible to choose a real $K_0$ and obtain a complex $G$, or, reciproquely, to choose a real $G$ and obtain a complex $K_0$ (cf. the discussion in \cite{J2G}). To that extend, the true situation is that one can choose a complex $K_0$ or a complex $G$ and obtain a corresponding $G$ or $K_0$. In fact, the relation $F(K_0^2,G^2)=0$ defines a multi-sheeted Riemann surface. An explicit example is given in fig. \ref{riem}.

The modes read as:
 $$
 H(X,Z,t)=\Re(u(X)e^{i\left(GZ-K_0 c T\right)})
 $$
 Therefore, whenever $G$ or $K_0$ is complex, this leads to an exponential damping in one direction of time or space and an exponential increasing in the other direction.
These modes are no longer tempered distribution in both time and space.
In the usual hilbertian framework of square integrable functions $L^2([0,H]\times \mathbb{R})$, there are no longer modes extending to infinity in an infinite lossy structure. A more convenient mathematical setting would be that of hilbertian triad, or Gel'fand triple \cite[chap. 1]{rigged}.
Still, the matching of two such modes can be bounded. Consider as a very simple example the functions:
$e^{-x}$ and $e^x$. By matching them at $x=0$, we obtain the function $e^{-|x|}$. This what happens with Laplace transform, which provides a decomposition over a half space: $f(x)=\int_0^{+\infty}F(p)e^{-px}dx$
which is convergent whenever $\Re(p)>0$.
We can apply the same kind representation in the lossy waveguide. Indeed, in such a structure the sources of the fiel cannot be placed at infinity as explained above. The sources here are of two kind: localized sources inside the waveguide such as quantum dots, atoms...and distributed sources inside the metal accounting for the fluctuations due to the absorption. These last can be accounted for by introducing Langevin current $\hat{J}$ such as in \cite{matloob}. The other can be obtained by using sums of Green function: the localized sources can be modelized as Dirac delta functions. When only the $X$ dependance is considered an infinite set of energies $K_{0n}$ are obtained at fixed $G$. When losses are introduced, the modes can be written in the form $\psi_n(X;\G)$ where $\psi_n(X;0)$ are the modes of the lossless guide. For a fixed $X$, these functions are all analytic in the variable $\G$, which implies that, as the family is complete at $\G=0$, it remains complete for a small enough $\G$ \cite{specg}. In a forthcoming work, we shall give explicit expressions for the Green function and apply it to the cavity quantum electrodynamics of plasmons.
\section{Conclusion}
We have studied the effect of confinement on the surface plasmons that can exist at an air-metal interface. We have given some approximate explicit solutions to the dispersion curves which has allowed us to demonstrate the strong coupling between the plasmon-like and the photon-like modes. We have shown the existence of strongly localized plasmon-like modes and also the onset of spatial dispersion when the filling ratio of metal is very low.
%\be 
%-\int \e^{-1} \d_x \psi_p \d_x \psi_n dx+ \int \left(k_{0p}^2 -\g_p^2 \e^{-1} \right)\psi_p \psi_n dx =0
% \ee
% \be 
%-\int \e^{-1} \d_x \psi_n  \d_x\psi_p dx+ \int \left(k_{0n}^2 -\g_n^2 \e^{-1} \right)\psi_n \psi_pdx =0
% \ee
% writing $\beta^2=k_{0}^2\e -\g^2 $, we obtain:
% $$
% \int \left(\beta_n^2-\beta_p^2\right) \e^{-1}\psi_p \psi_ndx =0
% $$
 
%% References
\bigskip

\noindent {\bf Acknowledgement} 

The financial support of the Institut Universitaire de France is gratefully acknowledged.

\bibliographystyle{elsarticle-num}

\begin{thebibliography}{00}

\bibitem{PRL} H. Iwase et al., ``Analysis of the Purcell effect in photonic and plasmonic crystals with losses,'' Opt. Express {\bf 18}, 16546 (2010).
\bibitem{bellessa} J. Bellessa et al., ``Strong Coupling between Surface Plasmons and Excitons in an Organic Semiconductor,'' Phys. Rev. Lett. {\bf 93}, 036404 (2004).
\bibitem{QD} K. Tanaka et al., ``Multifold Enhancement of Quantum Dot Luminescence in Plasmonic Metamaterials,'' Phys. Rev. Lett. {\bf 105}, 227403 (2010).
\bibitem{stockman}D. J. Bergman and M. I. Stockman, ``Surface Plasmon Amplification by Stimulated Emission of Radiation: Quantum Generation of Coherent Surface Plasmons in Nanosystems,'' Phys. Rev. Lett. {\bf 90}, 027402 (2003).
\bibitem{chalaev} E. Ozbay, ``Plasmonics: Merging Photonics and Electronics at Nanoscale Dimensions,'' Science {\bf 311}, 189-193 (2006).
\bibitem{chang} D. E. Chang et al., ``Quantum optics with surface plasmons,'' Phys. Rev. Lett. {\bf 97}, 053002 (2006).
\bibitem{Atwater} J. A. Dionne et al.,``Planar metal plasmon waveguides: frequency-dependent dispersion, propagation, localization, and loss beyond the free electron model,'' Phys. Rev. B {\bf 72}, 075405 (2005).
\bibitem{bra1} B. Guizal et al.,``Electromagnetic beam diffraction by a finite lamellar structure: an aperiodic coupled-wave method,'' J. Opt. Soc. Am. A  {\bf 20}, 2274-2280 (2003).
\bibitem{bra2} G. Granet and B. Guizal, ``Analysis of strip gratings using a parametric modal method by Fourier expansions,'' Opt. Comm. {\bf 255}, 1-11 (2005).
\bibitem{bra3} K. Edee et al., ``Beam implementation in a nonorthogonal coordinate system: Application to the scattering from random rough surfaces,'' J. Opt. Soc. Am. A  {\bf 25}, 796-804 (2008).
\bibitem{J2G} A. Archambault et al.,``Surface plasmon Fourier optics,'' Phys. Rev. B {\bf 79}, 195414 (2009).  
\bibitem{rigged} A. Bohm, H.-D. Doebner and P. Kielanowski (Eds.), Irreversibility and Causality, Springer, Berlin, 1998.
\bibitem{matloob} R. Matloob, R. Loudon, S. M. Barnett and J. Jeffers, ``Electromagnetic field quantization in absorbing dielectrics ,'' Phys. Rev. A {\bf 52}, 4823-4838 (1995).
\bibitem{specg} D. Felbacq, in preparation
\end{thebibliography}


 
\end{document}